\documentstyle[pre,multicol,aps,epsfig]{revtex}

\begin{document}

\draft

\tighten

\title{On the existence of internal modes of sine-Gordon kinks} 

\author{Niurka R.\ Quintero$^{*}$ and Angel S\'anchez$^{\dag}$} 

\address{Grupo Interdisciplinar de Sistemas
Complicados (GISC), Departamento de Matem\'aticas, 
Universidad Carlos III de Madrid,\\
Avenida de la Universidad 30, E-28911 Legan\'{e}s, Madrid, Spain}

\author{Franz G.\ Mertens$^{\ddag}$}

\address{Physikalisches Institut, Universit\"at Bayreuth,  
D-95440 Bayreuth, Germany}

\date{\today}

\maketitle

\begin{abstract} 
We study whether or not sine-Gordon kinks exhibit internal modes or 
``quasimodes.'' By considering the response of the kinks to ac forces
and initial distortions, we show that neither intrinsic
internal modes nor ``quasimodes'' exist in contrast to previous reports.
However, we do identify a
different kind of internal mode
bifurcating from the bottom edge of the phonon band 
which arises from the discretization of the system in the numerical
simulations, thus confirming 
recent predictions. 
\end{abstract}


\begin{multicols}{2}

\section{Introduction}
\label{intro}

Solitons, which were originally a concept arising in the study of integrable
systems \cite{scott}, have rapidly become a key paradigm in perturbed,
non-integrable
systems describing very many physical applications \cite{yurirev}. A very 
succesful picture of soliton dynamics under different perturbations has 
resulted from the use of collective coordinate (CC) techniques \cite{siam}: In many
instances, solitons behave basically like point-like particles, and therefore
their time evolution can be predicted by equations of motion which 
are ordinary differential equations.
Nevertheless, along the last decade it has been progressively realized that 
internal degrees of freedom of the solitons play a crucial role in a number of
problems: They govern resonant soliton (or solitary wave) collisions
\cite{David}, give rise to nontrivial soliton-impurity interactions 
\cite{Yura}, and can be excited both by ac forces \cite{prl} 
and by thermal noise \cite{grant}. Therefore, 
it is very important to assess whether or not physically relevant solitons 
possess internal modes. 

In this Rapid Communication we study solitons of 
the sine-Gordon (sG) equation, which arises in a diverse range of areas of
physics, covering from crystal dislocation theory to Josephson junctions 
\cite{scott,yurirev,siam}. In spite of the fact that linear stability 
analysis leads to only one eigenvalue in the discrete spectrum, 
corresponding to a zero frequency Goldstone mode \cite{yurirev}, 
Boesch and Willis \cite{willis} claimed that they found 
an internal ``quasimode'' above but close to the lower phonon band edge.
They described this ``quasimode'' as a long-lived oscillation of the 
width of the sG kink.
The possibility of such a mode had been suggested earlier by Rice 
\cite{Rice}, by means of a variational approach that reproduced with 
great accuracy the internal mode frequency of the $\phi^4$ model 
\cite{scott}
and predicted the existence of a similar mode for the sG problem.
To our knowledge, Ref.\ \cite{willis}
is the only paper reporting the observation of
this mode, although hints of its existence are scattered all over 
the soliton literature \cite{mas}. A few years later, it was 
theoretically found \cite{yuri}, within the Rice approach, that 
the internal mode of the sG and the $\phi^4$ kinks could be
excited by a constant external force or by an initial distortion;
however, no numerical evidence 
was presented to confirm these results. 
Therefore, we decided to revisit the
issue of the existence of the internal mode of the sG equation in order
to give a definitive answer to it. Our approach, based in part on our 
earlier findings \cite{prl},  overcomes the difficulty 
that the internal ``quasimode'' would lie within the phonon band by allowing 
us to look at driving frequencies well in the gap. 
As we will show below, this permits us to establish clearly that
this putative internal mode does not exist. Furthermore, we do observe
the excitation of a discreteness induced internal mode \cite{Braun,kivshar},
whose frequency is in excellent agreement with recent predictions by 
Kevrekidis and Jones \cite{jones}. We substantiate 
these claims by briefly recalling our main analytical results in Sec.\ 
\ref{colco}, collecting our numerical results in Sec.\ \ref{numso}, 
and discussing our conclusions in Sec.\ \ref{conclu}. 

\section{Resonances due to ac forces}
\label{colco}

We begin by considering the perturbed sG equation 
\begin{equation}
\begin{array}{c}
\phi_{tt} - \phi_{xx} = - \sin\phi +
\epsilon \sin(\delta t + \delta_{0}).  
\label{ecua1}
\end{array}
\end{equation}
where $\epsilon$, $\delta$, and $\delta_0$ are the amplitude, frequency,
and phase of the external periodic force. 
In \cite{prl} we analyzed analytically and numerically the above problem 
in the more general context of nonlinear (possibly damped)
 Klein-Gordon systems, 
which includes Eq.\ (\ref{ecua1}) as well as the $\phi^4$ model. 
Particularizing our results, obtained by means of the generalized 
travelling wave {\em Ansatz} \cite{franz} combined with the Rice {\em Ansatz} 
\cite{Rice}, we find that 
the momentum $P(t)$ and the width of the kink $l(t)$ 
(of a general nonlinear Klein-Gordon equation; specific constants for 
the sG case will be given below) obey the equations
\begin{eqnarray}
\frac{dP }{dt} &=& -  q\, \epsilon \sin(\delta t + \delta_{0}), \quad
P(t) \equiv \frac{M_{0} l_{0} \dot{X}}{l(t)},   
\label{ecua23}\\
\alpha \left(\dot{l}^{2} - 2 l \ddot{l} \right) &=&   
\frac{l^{2}}{l_{0}^{2}} \left (1 +  \frac{P^{2}}{M_{0}^{2}} \right) - 1,  
\label{ecua24}
\end{eqnarray}
where X(t) is the kink position. For
sG, $q=2\pi$, $M_{0}=8$, $l_{0}=1$ and $\alpha=\pi^2/12$. 
The equation for the momentum is linear and can be solved exactly, yielding 
\begin{equation}
P(t) =  \left[P(0)-
\frac{q\,\epsilon
\cos(\delta_{0})}{\delta} \right]+
 \frac{q\,\epsilon}{\delta} \,
\cos(\delta t + \delta_{0}) 
 \label{ecua4}
\end{equation} 
We thus see that the term $P(t)^2$ in Eq.\ (\ref{ecua24}) involves two 
frequencies, $\delta$ and $ 2 \delta$.
As shown in \cite{prl}, for the $\phi^4$ kink this leads to
dramatic resonance effects when the ac driving frequency $\delta$ is 
half the frequency $\Omega_i$ of the internal mode.
Without resonant excitation of kink modes, 
the kink oscillates around its initial position as a pure point-like 
particle, as shown analytically and numerically in \cite{EPJBus}.
However, as shown in \cite{prl} for the above parameters, in case 
an internal mode is excited as in the $\phi^4$ system, the system energy 
grows and eventually leads to chaotic motion of the kink.
In addition, the resonance at $\Omega_i/2$ turns 
out to be much stronger than that at $\Omega_i$, which 
makes it a convenient way to probe those modes:
In case the sG
kink would possess such a degree of freedom, we should observe similar
phenomena.
In \cite{willis}, 
the frequency of the numerically observed 
``quasimode'' was $\omega_s=1.004\pm 0.001$ in our
units, whereas Rice's theoretical
analysis leads to \cite{Rice} $\Omega_R=1.103$. 
In both cases, the corresponding half frequencies lie well within the
gap; we can thus force the system at the predicted resonances without
exciting much radiation which could mask any internal mode effects.

\section{Numerical results}
\label{numso}
  
In this Section, we present the results of our numerical search for 
internal modes or ``quasimodes'' of the sG equation.
We have computed the numerical solution of the perturbed 
sG equation (\ref{ecua1})
by using the Strauss-V\'azquez scheme \cite{stvaz},
the total length of the system being $L=100$, the steps being
$\Delta x=0.05$, $\Delta t=0.005$ (or other values when indicated).
Free boundary conditions were also used, and  
the final time in the simulations was $25 \, 000$, enough to detect any
possible resonance \cite{prl}. 

\subsection{ac force}
\label{acfor}

\begin{figure}
\epsfig{figure=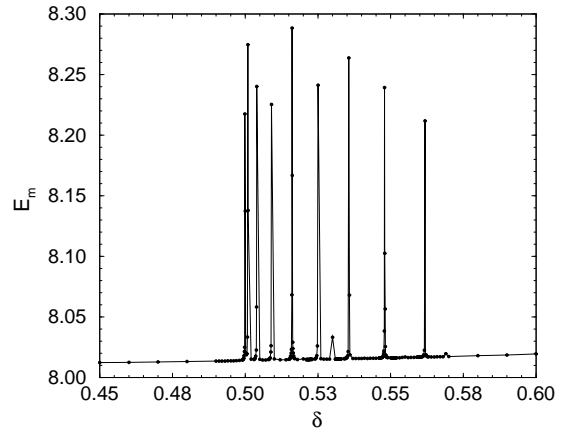, width=2.3in, angle=-90}
\caption{Mean energy of the 
sG system, in dimensionless units, when $\delta$ is close to $\Omega_{R}/2$. 
Points have been calculated from numerical simulations for 
$10\,000 \le t \le 25\,000$, with 
$\epsilon=0.01$, $\delta_{0}=\pi/2$, $u(0)=0$, $X(0)=0$. 
The parameters of the discretization are
$\Delta x=0.05$, $\Delta t=0.005$ and $L=100$.}
\label{f8}
\end{figure}

First, we have perturbed our system with an ac force of amplitude
$\epsilon = 0.01$ and phase $\delta_{0} = \pi/2$.
We have used a
sG kink at rest as initial condition. We 
have computed the energy (as defined in \cite{stvaz}) during the evolution,
and then the mean value of the energy in the time interval
$10\,000 \le t \le 25\,000$, studying a range of 
values of $\delta$ around half of the possible internal mode frequencies.
{}Figure \ref{f8} collects our results for the mean energy as a function 
of the frequency. We clearly see that the mean energy rapidly 
increases for 
some values of $\delta$ which do not coincide with any of the proposed 
internal modes or ``quasimodes.'' 
The first (leftmost) maximum lies below half the frequency of the
lower phonon band edge; therefore, this one can not be associated with
a phonon mode and we will discuss its origin in detail below. 
As for the other peaks, Table \ref{tab2.1} shows that 
we have been able to identify them with half of the values of the first
frequencies of the phonon band, $\omega_n$, which are given by 
\begin{equation}
\omega_{n}=\sqrt{1+\left[{(2 \pi n )}/{L}\right]^{2}},$
$n=1,2,3,\ldots, N=L/\Delta x.  
\label{fon}
\end{equation}
A comparison of 
the third and fourth columns of Table \ref{tab2.1},
where we list 
the half values of the frequencies corresponding to the first  
$8$ radiational modes and the numerical peak
frequencies ${\tilde{\omega}_{n}}$ in 
Fig.\ \ref{f8}, counting from the second maximum from the left, 
allows to verify that these values are nearly identical in all cases 
(see also the  relative differences
$|\omega_{n}/2-{\tilde{\omega}_{n}}|$ in Table \ref{tab2.1}). 
We stress that identification of radiation modes as the origin of
these resonances makes perfect physical sense: Indeed, an extended
CC approach including radiation modes leads to the conclusion that
the radiation modes should also be excited parametrically by ac 
drivings. 
In view of all this, the main conclusion we can
draw from Fig.\ \ref{f8} is the absence of 
any resonance due to 
the internal quasi-mode in the phonon band
reported by Boesch and Willis 
\cite{willis}, which, in view of the frequency they reported,
would lead to an extra peak located between the two rightmost
ones in Fig.\ \ref{f8}.  In this regard, it is important to note that
the accuracy of our simulations in detecting every radiation mode 
rules out the possibility that we have missed this quasi-mode
(at either $\omega_s$ or $\Omega_R$). 

Interestingly, we do observe an internal mode: Let us recall that the
first maximum of the energy occurs at
$\delta=\omega_{d}=0.4999$, i.e., $2 \omega_{d}=0.9998$ is below 
the lower phonon frequency $\omega_{ph}=1$ as we mentioned above. 
The reason for this peak is a 
novel internal mode 
of the sG kink, that appears as a bifurcation from the lowest phonon of 
the continuum equation, of frequency $\omega_{ph}=1$,
due to the effect of discreteness \cite{Braun,kivshar,jones}. 
As has been recently shown \cite{jones}, the frequency 
of this new internal mode is  
$\omega_{i} \approx \sqrt{1-(4/2025) (\Delta x)^4}$ (the $(\Delta x)^4$ 
dependence was 
already observed in \cite{Braun}). 
In our simulations, $\Delta x=0.05$, hence
$\omega_{i}=0.999998$, approximately equal to $2 \omega_{d}$. 
Therefore, the first peak in Fig.\ \ref{f8} corresponds to the parametric
resonance at half the value of the discreteness induced internal mode, 
very much like the resonance with the $\phi^4$ kink internal mode reported
earlier \cite{prl}. 

\subsection{Deformed initial kink}
\label{confo}

In view of our negative result about the internal quasi-mode of the sG kink,
we carried out more numerical simulations specifically designed to find it. 
Following \cite{willis}, we chose 
initial conditions given by 
\begin{equation}
\begin{array}{l}
{\displaystyle{
\phi(x,0)=4\,\mbox{arctan}\left(
\exp\left[\frac{x-X(0)}{l(0)}\right]\right)}},\\
\\
{\displaystyle{ 
\phi_t(x,0)=2\,\frac{\displaystyle{\left[
-\frac{u(0)}{l(0)}-\frac{x-X(0)}{l^2(0)}
\dot{l}(0)\right]}}{\displaystyle\mbox{cosh}\left[\frac{x-X(0)}{l(0)}\right]}
}}, 
\end{array}
\label{val-ini-sG}
\end{equation}
which correspond to a deformed kink if $l(0)\neq 1$ or $\dot{l}(0)\neq 0$.
By starting their simulations with 
such initial conditions, Boesch and Willis claimed that they were able
to excite the internal quasi-mode, and hence we hope that, if it is 
indeed present, we should find it. 

\begin{figure}
\epsfig{file=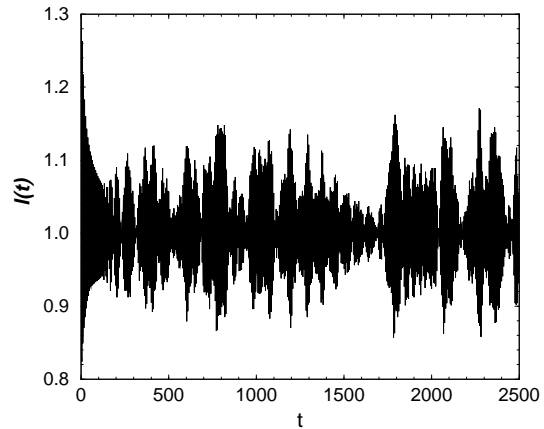,width=2.3in,angle=-90}
\caption{Evolution of the kink width, in dimensionless units, 
obtained from numerical simulations starting from a free, deformed kink, with 
$u(0)=0$, $l(0)=1$ and ${\dot{l}}(0)=0.3$. The parameters 
of the discretization are  $L=100$, $\Delta x=0.1$ and $\Delta t=0.01$.  
}
\label{f3.8}
\end{figure}

At this point, we have to recall the work of 
Majern\'\i kov\'a, Gaididei and Braun
\cite{yuri}, who, as mentioned in the introduction,
theoretically considered the problem of a deformed kink 
given by the expressions above driven by a external constant force. 
When applied to this problem, our 
CC approach yields the same 
two ordinary differential equations as in \cite{yuri} 
for $X(t)$ and $l(t)$, 
[Eqs.\ (\ref{ecua23})-(\ref{ecua24}) without the sine term, i.e., with 
only a constant force $\epsilon$]. 
We have solved these equations exactly for arbitrary initial conditions
\cite{usi} 
in terms 
of the Whittaker functions, concluding
that the kink width oscillates only if $l(0) \ne l_{s}$ 
(the natural width of the undistorted kink, $1/\sqrt{1-u(0)^2}$)
or ${\dot{l}}(0) \ne 0$ in (\ref{val-ini-sG}), whereas 
external force and dissipation (if present) only damp out
those oscillations. Hence, kink width oscillations 
should be easiest to detect without force and damping and, consequently,
here we consider only free propagation of an initially distorted kink
(a general analysis of the constant force problem will be presented 
elsewhere \cite{usi}). As a check of the validity of this procedure to 
probe internal modes, we
have numerically verified that the $\phi^4$ kink indeed
behaves in this fashion due to the excitation of the internal mode,
exhibiting oscillations in the kink width very close to those predicted
by the CC approach. We then turned to 
the sG case, and we found that the 
oscillations are not correctly described 
by the CC equations. In Fig.\ 
\ref{f3.8} we can see the evolution of the kink width computed from 
numerical simulation of the sG system, with $l(0)=1$, $u(0)=0$, and 
${\dot{l}}(0)=0.3$. It is clear that $l(t)$ is not a simple oscillatory 
function as in the $\phi^4$ case, and 
Fourier analysis yields 
the following frequencies:  
$\omega_{d}=0.9983$, ${\tilde{\omega}}_{1}=1.0034$, 
${\tilde{\omega}}_{2}=1.0083$, ${\tilde{\omega}}_{3}=1.0184$, 
${\tilde{\omega}}_{4}=1.0335$, ${\tilde{\omega}}_{5}=1.0511$, 
${\tilde{\omega}}_{6}=1.0712$, ${\tilde{\omega}}_{7}=1.0963$.
As before, all these frequencies are very close to the 
discreteness induced internal mode and the first few radiational
modes, with no evidence supporting the existence of the internal
quasi-mode above the phonon band edge.

\section{Conclusions}
\label{conclu}

The first conclusion of this work is that the existence of the 
proposed internal quasi-mode of the sG kink above the phonon 
band edge \cite{willis} is very unlikely. We have found 
neither resonances nor long-lived kink width oscillations
arising from such a mode when the kink is subject
to ac driving or to initial deformations, respectively. We are confident
that we have carefully explored all the range of relevant frequencies,
as we have accurately detected all possible resonances within that
range. In addition, the comparison with the phenomenology observed
for the $\phi^4$ kink, that possesses a true internal mode, shows 
that if the sG kink would have such an internal mode, we would have
detected its influence through numerical simulations as for the 
$\phi^4$ equation \cite{prl,usi}. In fact, now that we have 
seen that radiation modes are parametrically excited by their 
corresponding half-frequency drivings, we believe we can suggest an
explanation for the results of Boesch and Willis. 
Notice that they used Eq.\ (\ref{fon}) to calculate  
the lower phonon frequency, obtaining that   
$\omega_{1}=\sqrt{1+(2 \pi/1000)^2} = 1.00002$, whereas  
the internal ``quasimode'' frequency that they found 
in the numerical simulations was $\omega_{s}=1.004$. As those two 
values were largely different, they concluded that the quasi-mode was
well above the lowest phonon frequency and different from it. 
However, as we have discussed in the previous sections, 
they used initial conditions in which the initial kink 
velocity was not zero; in this case, it can be easily shown that 
the corresponding frequencies of the phonons are given by 
$$
{\displaystyle{
\bar{\omega}_{k}=\frac{\omega_{k}-k\,u(0)}{\sqrt{1-u^2(0)}}, \,\ \,\ 
\omega_{k}=\sqrt{1+k^2}.
}}
$$
If we now insert the parameters 
used in \cite{willis} to perform a numerical experiment similar to the 
one discussed here, we find 
$\omega_{1}=1.004$, i.e., the resonance observed by Boesch and Willis
took place in fact with the lowest frequency phonon {\em in the presence
of a moving kink}
and not with any internal quasi-mode.

The second conclusion of the present paper relates to the recently 
proposed internal modes induced by perturbations, in our case by 
discreteness \cite{Braun,kivshar,jones}. By the two different types of 
numerical simulations used in our research, we have detected the 
internal mode bifurcating from the edge of the phonon band 
(of the continuum problem). The frequency of that mode agrees qualitatively
very well with the one predicted in \cite{jones}. Therefore, our research 
fully confirms the results of those papers and, in addition, shows that 
these perturbation induced internal modes behave very much like intrinsic
ones.
Our result, along with the finding of similar 
internal modes in other nonintegrable sG systems \cite{nueva}, 
strongly supports the generality of the phenomenon predicted
in \cite{kivshar}.
Finally, the fact that we
are able to detect this novel mode reinforces our previous conclusion 
of the non-existence of the internal quasi-mode, as indeed the sG kink
responds to the ac driving or to deformations as it should if it had 
an internal mode, but for the discreteness induced one only. 
We thus believe that the present work definitely rules out the 
possibility of internal modes or ``quasimodes'' of sG kinks close to 
and above the lower phonon band edge.

\vspace*{-1cm}

\section{Acknowledgments}

We thank Yuri Kivshar and Yuri Gaididei for discussions. 
Work at GISC (Legan\'es) has been supported by 
DGESIC (Spain) grant PB96-0119. Travel between Bayreuth and Madrid has been
supported by ``Acciones Integradas Hispano-Alemanas'', a joint program of
DAAD (Az.\ 314-AI) and DGESIC.

\begin{table}[h]
\begin{center}
\begin{tabular}{|c|c|c|c|c|}
Mode($n$) &  $\omega_{n}$ & $\omega_{n}/2$ &
$\widetilde{\omega}_n$ & 
$\left|\omega_{n}/2-\widetilde{\omega}_n \right|$ 
\\
\hline
1 & 1.0019 & 0.5009 & 0.5009 & 8.6 10${}^{-5}$ \\
\hline
2 & 1.0079 & 0.5039 & 0.5040 & 7 10${}^{-5}$ \\
\hline
3 & 1.0176 & 0.5088 & 0.5090 & 1.9 10${}^{-4}$  \\
\hline
4 & 1.0311 & 0.5155 & 0.5161 & 5.5 10${}^{-4}$  \\
\hline
5 & 1.0482 & 0.5241 & 0.5250 & 9.1 10${}^{-4}$  \\
\hline
6 & 1.0687 & 0.5343 & 0.5356 & 1.3 10${}^{-3}$  \\
\hline
7 & 1.0924 & 0.5462 & 0.5479 & 1.7 10${}^{-3}$ \\
\hline
8 & 1.1192 & 0.5596 & 0.5618 & 2.2 10${}^{-3}$  
\end{tabular}
\end{center}
\caption{Comparison of the half of the first $8$ frequencies 
$\omega_{n}/2$ of 
radiational modes with the computed frequencies 
$\widetilde{\omega}_n$ 
from numerical simulations of (\ref{ecua1}) at which peaks in 
$E_{m}$ arise (see Fig.\ \ref{f8}).}
\label{tab2.1}
\end{table}

\end{multicols}

\end{document}